\documentclass[aps,prb,twocolumn,groupedaddress,showpacs]{revtex4}
\usepackage{graphicx}
\usepackage{bm}
\begin{document}
\title{Bulk Electronic structure of Na$_{0.35}$CoO$_{2}$.1.3H$_{2}$O}
\author{A. Chainani,$^{1}$ T. Yokoya,$^{2}$ Y. Takata,$^{1}$ K. Tamasaku,$^{3}$ M. Taguchi,$^{1}$ T. Shimojima,$^{2}$ N. Kamakura,$^{1}$ K. Horiba,$^{1}$ S. Tsuda,$^{2}$ S. Shin,$^{1,2}$
D. Miwa,$^{3}$  Y. Nishino,$^{3}$ T. Ishikawa,$^{3}$ M. Yabashi,$^{4}$  K. Kobayashi,$^{4}$ H. Namatame,$^{5}$ M. Taniguchi,$^{5}$
K. Takada,$^{6,7}$ T. Sasaki,$^{6,7}$ H. Sakurai,$^{6}$ and E. Takayama-Muromachi.$^{6}$}
\affiliation{$^{1}$Soft X-ray Spectroscopy Lab, RIKEN/SPring-8, 1-1-1 Kouto, Mikazuki-cho, Sayo-gun, Hyogo 679-5148, Japan}
\affiliation{$^{2}$Institute for Solid State Physics, University of Tokyo, Kashiwa,Chiba 277-8581, Japan}
\affiliation{$^{3}$Coherent X-ray Optics Lab, RIKEN/SPring-8, 1-1-1 Kouto, Mikazuki-cho, Sayo-gun, Hyogo 679-5148, Japan}
\affiliation{$^{4}$JASRI/SPring-8, 1-1-1 Kouto, Mikazuki-cho, Sayo-gun, Hyogo 679-5198, Japan}
\affiliation{$^{5}$HiSOR, Hiroshima University, 2-313 Kagamiyama, Higashi-Hiroshima 739-8526, Japan}
\affiliation{$^{6}$National Institute for Materials Science, Tsukuba, Ibaraki, 305-0044, Japan}
\affiliation{$^{7}$CREST, Japan Science and Technology Corporation}
\date{\today}
\begin{abstract}
High-energy (h$\nu$ = 5.95 KeV) synchrotron Photoemission spectroscopy (PES) is used to study bulk electronic structure of Na$_{0.35}$CoO$_{2}$.1.3H$_{2}$O, the layered superconductor. In contrast to 3-dimensional doped Co oxides, Co $\it{2p}$ core level spectra show well-separated Co$^{3+}$ and Co$^{4+}$ ions. Cluster calculations suggest low spin Co$^{3+}$ and Co$^{4+}$ character, and a moderate on-site Coulomb correlation energy U$_{dd}\sim$3-5.5 eV. Photon dependent valence band PES identifies Co $\it{3d}$ and O $\it{2p}$ derived states, in near agreement with band structure calculations.
\vspace{-0.3in}
\end{abstract}
\maketitle

The discovery of superconductivity in the hydrated Co oxide Na$_{0.35}$CoO$_{2}$.1.3H$_{2}$O is important in terms of a layered triangular lattice with superconductivity.\cite{Takada} Since it has two-dimensional CoO$_{2}$ layers consisting of (CoO$_{6}$) octahedra and Na$^{1+}$ content corresponds to Co$^{3+}$ valency in a matrix of Co$^{4+}$ ions, it is reminiscent of doping induced superconductivity as in high-T$_{c}$ cuprates. In spite of extensive investigations,\cite{Good} it has not been possible to achieve superconductivity in a 3-dimensional Co-oxide system. Electron-electron correlations between Co $\it{3d}$ electrons is believed to be substantial (on-site Coulomb energy, U$_{dd}$ = 3-5.5 eV) from electron spectroscopic studies,\cite{Elp,Chainani,Saitoh,Mizokawa,Parm} albeit less than copper oxides (U$_{dd}$ = 5-8 eV, Ref. 8). Theoretical studies,\cite{Baskaran,Kumar,Wang,Koshibae} including resonating-valence-bond models, predict fascinating properties for this system. Recent experiments signifying strong correlations show dimensional crossover\cite{Valla} and the relation of spin entropy with the large thermopower\cite{Cava} in the non-superconducting compositions.

From the point of conventional phonon-mediated superconductivity, 
weak or strong electron-phonon coupling leading to superconductivity also needs to be carefully investigated for the Co-oxide superconductors. This is because the doping dependent T$_{c}$'s are rather low,\cite{Schaak} with a maximum T$_{c}$ of $\sim$5 K. The layered Co oxides thus provide a new opportunity to study charge and spin dynamics in superconducting oxides. Recent studies indicate that Na$_{x}$CoO$_{2}$ is close to charge and spin ordering tendencies.\cite{Kunes, Terasaki} NMR studies\cite{Ray} on non-superconducting Na$_{x}$CoO$_{2}$ have concluded integral valent Co$^{3+}$ and Co$^{4+}$ ions reflecting charge order. Also, while intercalated water is necessary for superconductivity, its role in modifying the electronic structure is not yet clear. It is thus important to study the electronic structure of Na$_{x}$CoO$_{2}$.yH$_{2}$O as a function of Na content, x and water content, y.

Photoemission spectroscopy(PES) has provided a systematic enumeration of the  electronic structure of transition metal compounds\cite{Elp,Chainani,Saitoh,Mizokawa,Parm,Fujimori} based on the Zaanen-Sawatzky-Allen (ZSA) phase diagram.\cite{ZSA} Core-level PES provides valence states and a reasonable estimate of electronic structure parameters : on-site Coulomb energy (U$_{dd}$), charge-transfer(CT) energy($\Delta$) and hybridization strength(V). Further, while angle-resolved(AR) valence band(VB) PES is necessary to study experimental band structure, angle integrated VB-PES provides the transition
probability modulated density of states(DOS). The surface sensitivity of PES has often led to controversies regarding surface $\it{vis'-a-vis'}$ bulk electronic structure, and hence, high-energy(HE)-PES as well as site-selective PES are very important and promising.
\cite{Braico, Woicik} A recent development using a high-throughput fixed photon energy and a resolution of 240 meV at a kinetic energy of 5.95 KeV, is a valuable advance for investigating bulk electronic structure of materials.\cite{Kobayashi} Its efficacy was demonstrated for a high-K dielectric material for semiconductor applications.\cite{Kobayashi} The principal advantage of HE-PES is the high escape depth of emitted photoelectrons,\cite{NIST} enabling a truly bulk measurement. At h$\nu$ = 5.95 KeV, the escape depth for Co $\it{2p}$ and O $\it{1s}$ electrons is estimated to be $\sim$50 \AA, significantly higher than that with soft x-ray photons from a Mg- or Al-K$\alpha$ 
source($\sim$10 \AA).\cite{NIST} Since photo-ionization cross sections(PICS) become very low at high photon energies,\cite{Yeh} VB studies at h$\nu$ $\ge$ 5 keV were very difficult earlier, although the first core level study using 8 KeV photons was done nearly 30 years ago.\cite{Lindau}

We study VB and core-level HE-PES (h$\nu$ = 5.95 KeV) 
of the Co oxide superconductor, Na$_{0.35}$CoO$_{2}$.1.3H$_{2}$O and non-superconducting Na$_{0.7}$CoO$_{2}$.
Co $\it{2p}$ core level spectra show well defined Co$^{3+}$ and Co$^{4+}$ features in the normal phase of Na$_{0.35}$CoO$_{2}$.1.3H$_{2}$O. Cluster calculations indicate a moderate U$_{dd}\sim$3-5.5 eV. The O $\it{1s}$ spectrum of Na$_{0.35}$CoO$_{2}$.1.3H$_{2}$O shows a two-peak structure due to signals from CoO$_{2}$ layers and water, respectively. The VB spectra consisting of Co $\it{3d}$ and O $\it{2p}$ derived states are compared with soft x-ray (h$\nu$ = 700 eV) PES and reported band structure calculations(BSCs). The VB is similar for both compositions on the energy scale of the resolution used, suggesting important modifications only at a lower energy scale near the Fermi level(E$_{F}$),\cite{Valla} probably related to confined carriers in CoO$_{2}$ layers and/or a modified electron-phonon coupling. 

Polycrystalline samples of Na$_{0.35}$CoO$_{2}$.1.3H$_{2}$O 
and Na$_{0.7}$CoO$_{2}$ were made and characterized as described in Ref. 1. Magnetization measurements confirmed the bulk T$_{c}$ of 4.5 K for Na$_{0.35}$CoO$_{2}$.1.3H$_{2}$O. To ensure retention of water under vacuum conditions, freshly prepared samples were mounted on substrates with silver paste and also covered with it. After transferring samples to the measurement chamber and cooling to 100 K, they were scraped in-situ with a diamond file to obtain clean surfaces. The samples were then cooled to 15 K for HE-PES measurements, at a vacuum of 1 x 10$^{-10}$ Torr. HE-PES was performed at undulator beam line BL29XU, Spring-8(Ref. 26) using 5.95 KeV photons and a modified SES2002 electron analyzer. The energy width of incident x-rays was 70 meV, and the total energy resolution, $\Delta$E was set to $\sim$ 0.5 eV.   Soft x-ray PES (h$\nu$ = 700 eV) was performed at BL19B, KEK, PF, using a CLAM4 electron analyzer with  $\Delta$E $\sim$ 0.3 eV.  Samples were cooled to 30 K and the vacuum was 8 x 10$^{-10}$ Torr during measurements. Single crystal CoO was scraped in-situ and measured at 300 K to calibrate the energy scale. E$_{F}$ of gold was also measured to calibrate the energy scale.

\begin{figure}
\includegraphics[width={7.5cm}]{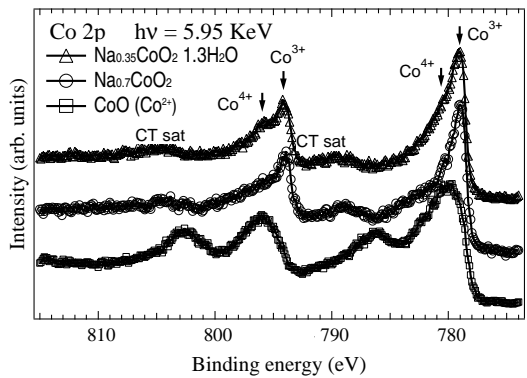}
\caption{\label{fig:epsart} Co 2p core level PES spectra of 
Na$_{0.35}$CoO$_{2}$.1.3H$_{2}$O, Na$_{0.7}$CoO$_{2}$ and CoO obtained using h$\nu$ = 5.95 KeV photons.}
\vspace{-0.25in}
\end{figure}
\begin{figure}
\includegraphics[width={7.5cm}]{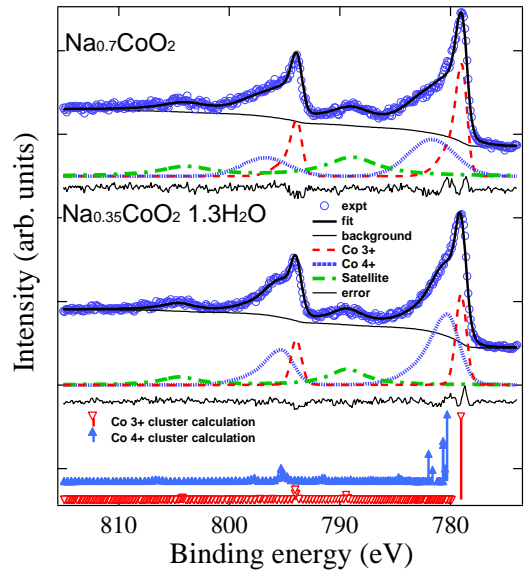}
\caption{\label{fig:epsart}(Color online) A least-squares curve-fit to the Co 2p spectrum of Na$_{0.35}$CoO$_{2}$.1.3H$_{2}$O and Na$_{0.7}$CoO$_{2}$, shows contributions of Co$^{3+}$ and Co$^{4+}$ states in the 2p$_{3/2}$ and 2p$_{1/2}$ main peaks, with a single peak assumed for satellites. Line diagrams show calculated low-spin Co$^{3+}$ and Co$^{4+}$ states.}
\vspace{-0.25in}
\end{figure}
Figure 1 shows the Co $\it{2p}$ core level PES spectra 
of Na$_{0.35}$CoO$_{2}$.1.3H$_{2}$O, Na$_{0.7}$CoO$_{2}$ and CoO obtained using h$\nu$ = 5.95 KeV photons. The Co 2p spectrum of Na$_{0.35}$CoO$_{2}$.1.3H$_{2}$O exhibits main peaks derived from Co $\it{2p}$$_{3/2}$ and $\it{2p}$$_{1/2}$ due to  spin-orbit splitting, and two small humps or satellites at $\sim$10 eV higher binding energy(BE) from the main peaks.  The $\it{2p}$$_{3/2}$ peak itself consists of two peaks. This is clear in the $\it{2p}$$_{1/2}$ region with well-separated peaks in raw spectra. A simple interpretation is that the low BE peak is due to Co$^{3+}$ and the high BE peak is due to Co$^{4+}$ states. In Fig. 2 we overlay a least-squares curve fit on the data of Na$_{0.35}$CoO$_{2}$.1.3H$_{2}$O and Na$_{0.7}$CoO$_{2}$, obtained using asymmetric Voigt functions and a Shirley background. The fits resolve contributions of Co$^{3+}$ and Co$^{4+}$ features in the main peaks, as seen in the decomposition. We used a single feature for the satellites for simplicity. 
The peak widths for Co$^{4+}$ feature and satellite in Na$_{0.7}$CoO$_{2}$ are broader than in Na$_{0.35}$CoO$_{2}$.1.3H$_{2}$O, suggesting larger inhomogeneity in valency  due to increased Na doping. The crystal structure analysis of both samples indicated absence of impurity phases within experimental accuracy. Further, we have carefully checked that
BE's for Co$^{3+}$  main peaks are actually lower than the corresponding peaks of Co$^{2+}$ in CoO,  e.g. $\it{2p}$$_{3/2}$ peak is at 779.0 eV and 780.0 eV, respectively (Fig. 1). The CoO spectra match those obtained with a MgK$\alpha$ source extremely well in BE and spectral shapes.\cite{Elp,Parm} Well-separated core-level features are observed in classic non-oxide charge-density wave(CDW) systems\cite{CDW} as well as intermediate valence materials without static charge order\cite{Steeneken}, because PES is a fast probe. 
Earlier work on 3-dimensional perovskite oxides La$_{1-x}$Sr$_{x}$CoO$_{3}$ (x = 0.0-0.4) showed essentially a single peak at the same BE, and no clear separation into Co$^{3+}$ and Co$^{4+}$states.\cite{Chainani,Saitoh} The peak width broadened initially with doping for x = 0.1, but across the semiconductor-metal transition at x = 0.2 in La$_{1-x}$Sr$_{x}$CoO$_{3}$, the peaks became narrower for increasing x due to uniform non-integral valency at Co-sites.\cite{Chainani} In misfit layered Co-oxides (BiPb)Sr-Co-O with an average valence of 3.33 and 3.52, no clear Co$^{4+}$ separated from Co$^{3+}$ was concluded.\cite{Mizokawa} Surprisingly, for oxide systems which show charge-ordering or disproportionation, such as Pr$_{0.5}$Sr$_{0.5}$MnO$_{3}$, perovskite ferrites, etc. well-separated integral valence features are not observed.\cite{Kurmaev,Bocquet} It is due to the ground state being dominated by a CT d$^{n+1}$L$^{1}$ (L is a ligand hole) rather than a d$^{n}$ configuration, based on model Hamiltonian cluster calculations.\cite{Bocquet} From similar cluster calculations (details are described in Ref. 30 and results of one such calculation for low spin Co$^{3+}$ and Co$^{4+}$ are shown as line diagrams in Fig. 2), we obtain the electronic structure parameters of U$_{dd}$ = 5.5 eV, $\Delta$ = 4.0 eV and V = 3.1 $\pm$ 0.2 eV which describe the Co $\it{2p}$ spectral features fairly well. For simplicity, we use the same parameter values for Co$^{3+}$ and Co$^{4+}$ except for crystal field splitting, 10Dq. The 10Dq values for Co$^{3+}$ and Co$^{4+}$ are 2.5 eV and 4.0 eV, respectively. The uncertainty in $\Delta$ is $\pm$ 0.5 eV while the calculated spectra were very similar for U$_{dd}$ = 3.0-5.5 eV, consistent with earlier work.\cite{Elp,Chainani,Saitoh,Mizokawa,Parm} The ground state character for Co$^{3+}$  is $\it{3d}$$^{6}$ = 57.0 \%, $\it{3d}$$^{7}$L$^{1}$ = 38.1 \% and 
$\it{3d}$$^{8}$L$^{2}$ = 4.9 \%, while that for Co$^{4+}$ is 
$\it{3d}$$^{5}$ = 57.4 \%, $\it{3d}$$^{6}$L$^{1}$ = 37.6 \% and $\it{3d}$$^{7}$L$^{2}$ = 5.0 \%. We have also checked for high-spin Co$^{3+}$ and Co$^{4+}$ configurations but the results are not compatible with the data. The analysis suggests that Na$_{0.35}$CoO$_{2}$.1.3H$_{2}$O and Na$_{0.7}$CoO$_{2}$ contain low spin Co$^{3+}$ and Co$^{4+}$ configurations, consistent with magnetic measurements.\cite{Terasaki} The calculations also show CT character of the satellites. The results indicate an electronic structure of mixed character, but more Mott-Hubbard-like rather than CT-like in terms of the ZSA phase diagram.  

\begin{figure}
\includegraphics[width={7.5cm}]{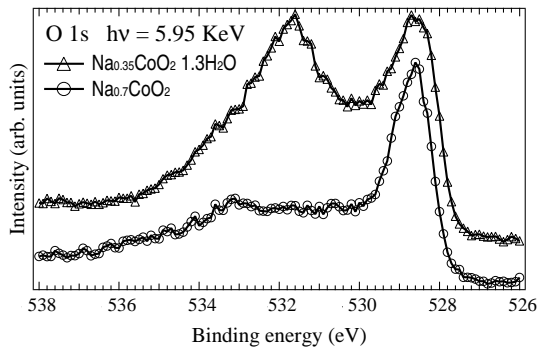}
\caption{\label{fig:epsart}O $\it{1s}$ core level PES spectra of Na$_{0.35}$CoO$_{2}$.1.3H$_{2}$O and Na$_{0.7}$CoO$_{2}$ obtained using h$\nu$ = 5.95 KeV photons, showing the presence of water derived O $\it{1s}$ signal in superconducting composition.}\vspace{-0.25in}
\end{figure}

While the main peaks of Co$^{3+}$ and Co$^{4+}$ features are well-separated, it is
clear from the cluster calculations that Co $\it{2p}$ spectra consist of 
degenerate multiple features at higher BEs. The satellite intensity is also large, being $\sim$70 \% of the Co$^{3+}$ main peak. It is hence difficult to estimate the actual Co$^{3+}$ : Co$^{4+}$ relative concentrations although the main peak intensities are roughly consistent (within 10 \%) of the nominal concentrations. The present studies are consistent with integral valent charge order measured by NMR studies.\cite{Ray} Although LDA + U calculations\cite{Kunes} for Na$_{0.33}$CoO$_{2}$ indicate a correlation driven charge order with a ferromagnetic ground state, the absence of ferromagnetic order and suppression of Co moments on introducing water in Na$_{0.35}$CoO$_{2}$.1.3H$_{2}$O(Ref. 32) suggests an additional input to the electronic structure, most likely strong electron-phonon coupling as in regular CDW transitions.

The O $\it{1s}$ core level HE-PES spectra of Na$_{0.35}$CoO$_{2}$.1.3H$_{2}$O and Na$_{0.7}$CoO$_{2}$ are shown in Fig. 3.  The spectrum of Na$_{0.35}$CoO$_{2}$.1.3H$_{2}$O has two peaks at BEs of 528.6 eV and 531.6 eV.  The peak at 528.6 eV is the oxygen 1s core level from the CoO$_{2}$ layers, as has been observed in layered Co oxides.\cite{Mizokawa}  The peak at 531.6 eV is ascribed to O $\it{1s}$ core level of water, as is evident from its BE.\cite{Qiu} While the 528.6 eV peak is present in Na$_{0.7}$CoO$_{2}$, the high BE feature at 531.6 eV is missing. A weak intensity feature at a still higher BE of 533 eV is observed, possibly due to carbonate-like contamination which is below the detection limit of x-ray diffraction. The results show that the superconducting sample contains water, which is absent in the nonsuperconducting compound, as expected from the compositions.

\begin{figure}
\includegraphics[width={7.5cm}]{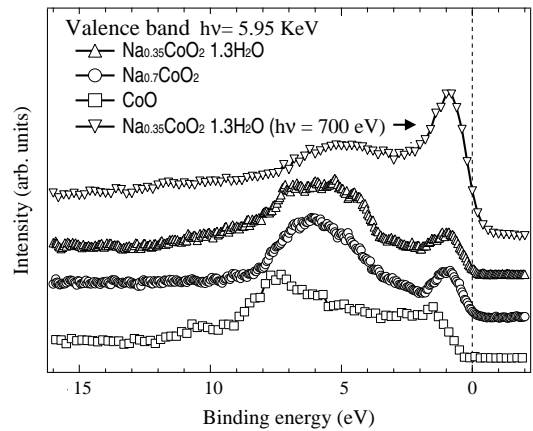}
\caption{\label{fig:epsart}Valence band HE-PES (h$\nu$ = 5.95 KeV) spectra of Na$_{0.35}$CoO$_{2}$.1.3H$_{2}$O, Na$_{0.7}$CoO$_{2}$ and CoO, and of Na$_{0.35}$CoO$_{2}$.1.3H$_{2}$O obtained using h$\nu$ = 700 eV photons. The Co 3d states are enhanced in the h$\nu$ = 700 eV spectrum.}
\vspace{-0.25in}
\end{figure}
The HE-PES VB spectra of Na$_{0.35}$CoO$_{2}$.1.3H$_{2}$O, Na$_{0.7}$CoO$_{2}$ and CoO are shown in Fig. 4, along with the soft x-ray (h$\nu$ = 700 eV) spectrum of Na$_{0.35}$CoO$_{2}$.1.3H$_{2}$O.  For Na$_{0.35}$CoO$_{2}$.1.3H$_{2}$O, the HE-PES spectrum shows a low intensity peak at 0.9 eV and a broad structure centered around 6 eV. In comparison, the h$\nu$ = 700 eV spectrum shows a higher intensity 0.9 eV peak with the leading edge crossing E$_{F}$ and a broad peak centered at 5 eV(data normalized at 5 eV BE). The spectral changes for the two photon energies arise from changes in PICS. This is confirmed by comparing the present HE-PES CoO data with that reported in ref. 34 for h$\nu$ = 600 eV. The spectral intensity changes and comparison with BSCs(Ref. 35) indicate that the feature at 0.9 eV is due to Co $\it{3d}$ states and the broad peak centered at 5-6 eV is dominated by O $\it{2p}$ states. The observed relative intensity changes indicate deviations from available calculated atomic PICS at h$\nu$ = 8.0 KeV\cite{Yeh} which suggest higher relative intensity of the Co $\it{3d}$ states compared to O $\it{2p}$ states.  BSCs for Na$_{0.5}$CoO$_{2}$ indicate a peak closer to E$_{F}$, with high DOS at E$_{F}$ derived from Co $\it{3d}$ t$_{2g}$ states, which is separated from the Co $\it{3d}$ e$_{g}$ states located in the unoccupied states.\cite{Singh} While oxides can show a contamination peak around 10 eV, the feature at 10.5 eV in CoO is intrinsic as it is observed for cleaved single crystals\cite{Parm,Ghir} and in the present case. It is clearly absent in Na$_{0.7}$CoO$_{2}$. A weak tailing feature between 9-12 eV is observed in Na$_{0.35}$CoO$_{2}$.1.3H$_{2}$O. Comparing with studies on interaction of water with a high-T$_{c}$ cuprate\cite{Qiu} and its absence in Na$_{0.7}$CoO$_{2}$, we attribute it to the water present in Na$_{0.35}$CoO$_{2}$.1.3H$_{2}$O. But for this feature, the Co $\it{3d}$ and O $\it{2p}$ derived states are similar in Na$_{0.35}$CoO$_{2}$.1.3H$_{2}$O and Na$_{0.7}$CoO$_{2}$. The VB of CoO shows a feature at nearly 1.6 eV consisting of Co $\it{3d}$ states and a higher BE broad feature due to O $\it{2p}$ states at about 7 eV. A comparison indicates that in Na$_{0.35}$CoO$_{2}$.1.3H$_{2}$O and Na$_{0.7}$CoO$_{2}$, the Co $\it{3d}$ feature is shifted to lower BE compared to CoO, as in Co $\it{2p}$ core levels(Fig. 1). 

Recent high-resolution ARPES studies on nonsuperconducting Na$_{x}$CoO$_{2}$(x = 0.5 - 0.7) also suggest consistency with BSCs, but with a renormalization of electronic states on a low energy scale of ~100 meV.\cite{Valla,Hasan} This energy scale is beyond present HE-PES measurements. A more accurate analysis at and very near E$_{F}$ in superconducting Na$_{0.35}$CoO$_{2}$.1.3H$_{2}$O requires higher resolution measurements, preferably with ARPES, to obtain the energy and momentum resolved electronic structure. The interplay of electron-electron correlations and strong electron-phonon coupling\cite{Rosch} of renormalized carriers in Na$_{0.35}$CoO$_{2}$.1.3H$_{2}$O could stabilize a `composite glue' for pairing, driven by a change in hybridization or intersite Coulomb interactions.

In conclusion, HE-PES provides normal state bulk electronic structure of Na$_{0.35}$CoO$_{2}$.1.3H$_{2}$O. In contrast to 3-dimensional doped Co oxides, the Co $\it{2p}$ core level spectra show well-separated Co$^{3+}$ and Co$^{4+}$ ions. Cluster calculations suggest low spin Co$^{3+}$ and Co$^{4+}$ states, and a moderate on-site U$_{dd}$ $\sim$3-5.5 eV. Valence band PES identifies Co $\it{3d}$ and O $\it{2p}$ derived states, nearly in agreement with BSCs. 
		
AC thanks Professor O. Gunnarsson for very valuable discussions. We thank Drs. M. Arita, T. Tokushima and K. Shimada for valuable experimental support.

\end{document}